\documentstyle[aps,multicol,prl,epsfig]{revtex}

\begin{document}

\title{On the modulational stability of Gross-Pitaevskii type equations in 1+1 dimensions}

\author{G. Theocharis$^1$, Z. Rapti$^2$, P.G. Kevrekidis$^2$, 
D.J. Frantzeskakis$^1$ and V.V. Konotop$^3$}
\address{ $^{1}$
Department of Physics, University of Athens,
Panepistimiopolis, Zografos, Athens 15784, Greece \\
$^2$ Department of Mathematics and Statistics, 
University of Massachusetts, Amherst MA 01003-4515, USA \\
$^3$ Centro de F\'{\i}sica 
Te\'orica e Computacional, Universidade de Lisboa, Av. Prof. Gama Pinto, 2, Lisboa 1649-003, Portugal}

\maketitle

\begin{abstract}

The modulational stability of the nonlinear Schr{\"o}dinger (NLS) equation 
is examined in the case with  a quadratic external potential. This study is 
motivated by recent experimental studies in the context of matter waves in 
Bose-Einstein condensates (BECs). The theoretical analysis invokes a lens-type 
transformation that converts the Gross-Pitaevskii into a regular NLS equation 
with an additional growth term. This analysis suggests the particular 
interest of a
specific time-varying potential ($\sim (t+t^{\star})^{-2}$). We examine both
this potential, as well as the time independent one numerically and
conclude by suggesting experiments for the production of solitonic 
wave-trains in BEC.

\end{abstract}

\vspace{2mm}

\section{Introduction}

Intensive studies of Bose-Einstein condensates (BECs) \cite{review} have 
drawn much attention to nonlinear excitations in them. Recent experiments 
have achieved to generate  topological structures, such as vortices 
\cite{vortex} and vortex lattices \cite{vl}, as well as solitons. 
Especially, as far as the latter are concerned, two types of solitons have 
been created, namely {\it dark} solitons \cite {dark,dexp2,nsbec} for 
condensates with repulsive interactions and {\it bright} ones \cite{bright} 
for condensates with attractive interactions. Dark solitons are density dips 
characterized by a phase jump of the wavefunction at the position of the dip, 
and, thus, can be generated  by means of 
phase-engineering techniques. 
Bright solitons, which were only recently created in BECs of $^7$ Li, are 
characterized by a localized maximum in the density profile without any 
phase jump across it. In the relevant experiments, this type of solitons was 
formed upon utilizing a Feshbach resonance to change the sign of the 
scattering length from positive to negative.
An interesting question concerns how such solitary wave structures may 
arise (i.e., which is the underlying physical mechanism for their 
manifestation and how they may be generated) in this novel context of 
matter waves in BECs. 

It is well-known that the dynamics of the BEC wavefunction is described (at the
mean field level, which is an increasingly accurate description, as the 
zero temperature limit is approached) by the Gross-Pitaevskii (GP) equation, 
a variant of the well-known
nonlinear Schr{\"o}dinger (NLS) equation \cite{sulem}, which incorporates 
an external trapping potential term. In the context of the ``traditional'' 
NLS equation (without the external potential), perhaps the most standard 
mechanism through which bright solitons and solitary wave structures appear 
is through the activation of the modulational instability (MI) of plane 
waves: In this case, the continuous wave (cw) solution of the NLS equation 
becomes unstable towards the generation of a chain of bright solitons. 
It is the purpose of this work to demonstrate that, under certain conditions, 
this may also happen in the case of the GP equation as well.

The MI is a general feature of continuum  as well as discrete nonlinear wave 
equations and its demonstrations span a diverse set of disciplines, ranging 
from  fluid dynamics \cite{benjamin67} (where it is usually referred to as the 
Benjamin-Feir instability) and nonlinear optics \cite{ostrovskii69} to 
plasma physics \cite{taniuti68}. Additionally, the MI has been examined 
recently in the context of optical  lattices in BECs both in one-dimensional 
and quasi-one dimensional
systems, as well as in multiple dimensions. In such settings, 
it has been predicted theoretically \cite{konotop,smerzi}
and verified experimentally \cite{cattal,kasevich} to 
lead to destabilization of plane waves, and in turn to
delocalization in momentum space (equivalent to localization in position
space, and the formation of solitary wave structures).

In this paper, we discuss the MI conditions for the continuous NLS equation 
in (1+1)-dimensions (1 spatial and 1 temporal)
\begin{eqnarray}
i u_t + u_{xx} + s |u|^2 u + V(x) u=0,
\label{req1}
\end{eqnarray}
in the presence of the external potential $V(x)$. This equation is actually 
a dimensionless effective GP equation, which describes the evolution of the 
wave-function of a quasi-one-dimensional (either pancake or cigar-shaped) BEC. In this 
context, we will consider the harmonic
potential \begin{eqnarray}
V(x)=-k(t) x^2,
\label{req3}
\end{eqnarray}
which is relevant, in particular, to experimental setups in which 
the (magnetic) trap is strongly confined in the 2 directions, while it is
 much shallower in the third one \cite{review}. The prefactor $k(t)$ is 
typically fixed in current experiments, but adiabatic changes in the 
strength (and, in fact, even in the location of the center) 
of the trap are experimentally feasible, hence we examine the more general
time-dependent case.

A self-consistent reduction of a 3D GP equation to a 1D NLS equation with 
external potential can be provided by means of the multiple-scale expansion. 
In the case of a cigar-shaped BEC such an expansion exploits the small 
parameter $\epsilon^2=8\pi N a_s a_\bot/a_0^2\ll 1$, where $N$ is a 
number of atoms, $a_s$ is the s-wave scattering length, 
$a_{\perp}=\sqrt{\hbar/m \omega_{\perp}}$ and 
$a_{0}=\sqrt{\hbar/m \omega_0}$ are the transverse and longitudinal 
harmonic oscillator lengths, $\omega_{\perp}$ and $\omega_{0}$ are the
 harmonic frequencies corresponding to the strong confinement cross-section
 and to the cigar axis, and $m$ is the atomic mass 
(see e.g., \cite{konotop} for details). In the present paper $a_0$ will be a 
varying quantity, and then in the estimates $a_0$ should be understood as an 
effective averaged quantity.  Smallness of the parameter $\epsilon$ means 
weakness of the two-body interactions compared with the kinetic energy of 
the atoms. Then 
the complex field $u$ in Eq. (\ref{req1}) represents the rescaled mean-field
wavefunction of the condensate according to
\begin{eqnarray}
u(x,t)= \left(\frac{8 \pi |a_s| \hbar}{m \omega_{\perp}} \right)^{1/2} \Psi,
\label{add1}
\end{eqnarray}
where $\Psi$ is the original order parameter. In this rescaling of the 
GP (resulting in Eq. (\ref{req1})),
$x$ is normalized to the harmonic oscillator length $a_\bot$,
time is normalized to the corresponding oscillation 
period, 
and the potential $V(x)$ is measured in units of $\hbar^2 a_{\bot}^{2}/8m$. 
The transverse distribution of the order parameter has been taken into account.
Notice also that in Eq. (\ref{req1}) the subscripts 
denote partial derivatives with respect to the index, while 
$s=-\mbox{sign}(a_s)
\in \{1,-1\}$ illustrates the focusing ($+1$) or defocusing ($-1$)
nature of the nonlinearity (which represents the attractive or
repulsive nature of the inter-atomic interactions respectively \cite{review}).

After briefly reviewing (in section II) the MI for the NLS case, we 
proceed to our
main aim which is to study this instability in the context of the
GP of Eq. (\ref{req1}), with the potential of Eq. (\ref{req3}).
In section III, we show that a lens-type transformation, which transforms 
the GP equation into a relatively simpler form of the NLS equation, 
provides insight
in the latter case. Two interesting cases are singled out:
the case where $k(t) \equiv k$ (i.e., for a fixed trap) and the
case of $k(t) \sim (t+t^{\star})^{-2}$, which naturally arises in this
setting. In section IV, we investigate these cases numerically and find
a variety of interesting results including the generation of 
solitary wave trains. This result indicates that the MI is indeed an 
underlying physical mechanism explaining the formation of matter-wave 
soliton trains. Finally, we conclude with the discussion of section V 
which suggests this method as an experimental technique
for the generation of soliton patterns in BEC.

\section{Modulational Instability for NLS}

We start by recalling the results for the modulational stability of
the NLS (\ref{req1}) without an external potential, i.e. for $V(x)\equiv 0$:
\begin{equation}
iu_{t}+u_{xx}+ s |u|^2 u=0               
\label{req4}
\end{equation}
We look for perturbed plane wave solutions of the form
\begin{equation}
u(x,t)=(\phi+\epsilon b) \exp[i ((q x-\omega t) +\epsilon \psi(x,t))]
\label{req5}
\end{equation}
analyzing the $O(\epsilon)$ terms as
\begin{eqnarray}
b(x,t)=b_{0} \exp(i\beta(x,t)), \quad
\psi(x,t)=\psi_{0} \exp(i\beta(x,t)).
\label{req6}
\end{eqnarray} 
Using
$
\beta(x,t)=Q x-\Omega t,
$
the dispersion relation connecting the  wavenumber $Q$ and
frequency $\Omega$ of the perturbation (see e.g., \cite{sulem})
is found to be of the form
\begin{equation}
(-\Omega+2 q Q)^{2}=Q^{2} (Q^{2}-2 s \phi^{2})
\label{req8}
\end{equation}

This implies that the instability region for the NLS in the absence
of an external potential, appears for perturbation wavenumbers 
$Q^2<{2 } \phi^2$, 
and in particular {\it only for focusing nonlinearities}
(to which we will restrict our study from this point onwards).

\begin{figure}
\centering
{\epsfig{file=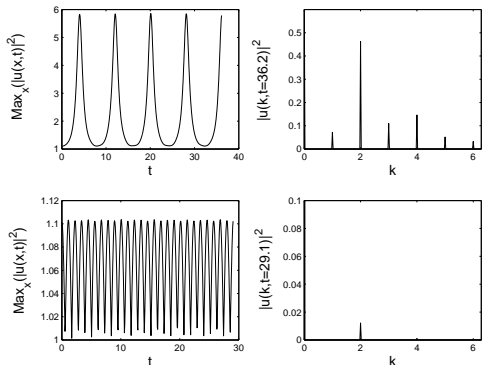, width=6.4cm,angle=0, clip=}}
\caption{The evolution of the maximum amplitude 
(left panels) and the Fourier transform at the ending
time of the simulation (right panels) for a modulationally
unstable case $Q=1$ (top panels) and a modulationally
unstable case $Q=2$ (bottom panels). An initial perturbation
of $0.05 \sin(Q x)$ was added to the uniform solution of 
$\phi=1$.}
\label{rfig1}
\end{figure}

A natural question is how this instability is manifested for
wavenumbers which satisfy the above condition. An example
is shown in Fig. \ref{rfig1}.

There are two principal ways in which the instability can be detected
(see Fig. \ref{rfig1}). One of them is by probing the maxima of the
original plane wave (notice that to avoid problems with the boundaries
the simulation shown in the figure was performed with periodic boundary
conditions). In the modulationally unstable case, we have periodic 
recurrence of structures with very large amplitude, while in the 
modulationally stable case of $Q=2$, the perturbation only causes
small amplitude oscillations. In the Fourier picture,
the unstable perturbation generates side bands of higher harmonics
as is well-known \cite{book}, while similar structures are absent in 
the modulationally stable case.

\section{Modulational Instability for NLS with Quadratic Potential}

The quadratic potential of Eq. (\ref{req3}) is clearly the most
physically relevant example of an external potential in the BEC
case, given the harmonic confinement of the atoms by the
experimentally used magnetic traps \cite{review}. 

To examine the MI related properties in this case,
we use a lens-type transformation \cite{sulem} 
of the form:
\begin{equation}
u(x,t)=\ell^{-1} \exp(i f(t) x^2)v(\zeta,\tau)
\label{req11}
\end{equation}
where $f(t)$ is a real function of time, $\zeta=x/\ell(t)$ and $\tau=\tau(t)$. 
To preserve the scaling we choose \cite{sulem,skk}
\begin{eqnarray}
\tau_{t}=1/\ell^2
\label{req13b}
\end{eqnarray}
The resulting equations can be satisfied by demanding that:
\begin{eqnarray}
f_{t}=-4f^2-k(t)
\label{req14a}
\\
\ell_{\tau}/\ell=4 f \ell^2.
\label{req14b}
\end{eqnarray}
Taking into account (\ref{req13b}) the last equation can be solved:
\begin{equation}
\label{req14c}
\ell(t)=\ell(0)\exp\left(4\int_0^tf(s)ds\right).
\end{equation}
This problem of finding the time dependence of the parameters is then
reduced to the solution of Eq. (\ref{req14a}).

Upon the above conditions, the equation for $v(\zeta,\tau)$ becomes
\begin{equation}
iv_{\tau}+v_{\zeta\zeta}+|v|^2 v-2 i\lambda v =0,  
\label{req15}
\end{equation}
where 
\begin{equation}
f \ell^2=\lambda,
\label{req16}
\end{equation}
and generically $\lambda$ is real and depends on time. Thus we retrieve 
NLS with an additional term, which represents
either  growth (if $\lambda>0$) or dissipation (if $\lambda<0$). 
 


A particularly interesting case is that of $\lambda$ constant. 
Then from the system of equations
(\ref{req13b})-(\ref{req14b})
and (\ref{req15}) 
it follows that $k$ must have a specific form. $f$,$\ell$ and 
$\tau$ can then be determined accordingly. 
In fact, the system (\ref{req13b})-(\ref{req14b})
and (\ref{req15})  with $\lambda$ constant has as its
solution
\begin{eqnarray}
k(t)&=&(t+t^{*})^{-2}/16
\label{req18a}
\\
f(t)&=&(t+t^{*})^{-1}/8
\label{req18b}
\\
\ell(t)&=&2 \sqrt{2 \lambda} \sqrt{t+t^{*}}
\label{req18c}
\\
\tau(t)&=&\ln (\frac{t+t^{*}}{t^{*}})/8 \lambda.
\label{req18d}
\end{eqnarray}
Notice that per the assumption of 
an imaginary phase in the exponential of Eq. (\ref{req11}), that
our considerations are valid only for $\lambda \in R$. 
In the above equations $t^{\star}$ is an arbitrary constant whose sign is related to the sign of $\lambda$; $t^\star\lambda>0$ and which essentially 
determines the ``width'' of the trap at time $t=0$ according 
to Eq. (\ref{req18a}). Notice that $t^\star <0$  
(i.e. the case of dissipation in Eq. (\ref{req15})) describes a BEC in a 
shrinking trap,
while the case $t^\star >0$  (i.e. the case of growth in Eq. (\ref{req15}))) 
corresponds to a broadening condensate.

In this case the  modulational condition remains unchanged, but now $\omega$ satisfies the dispersion  relation $\omega=q^{2}-\phi^{2}+2 \imath \lambda$, so the growth 
(if $\lambda>0$) or 
dissipation (if $\lambda<0$) is inherent in 
equation (\ref{req5}). Moreover, all the terms are 
modulated by the constant growth (or decay) rate $\exp(2 \lambda \tau)$, and 
the instability (when present) will be developing according to the form 
$v\sim \exp \left(i (Q \zeta-\Omega_r \tau)  + (\nu+2 \lambda) \tau\right)$ 
with $\Omega=\Omega_r + i \nu$. $\Omega_r=2 q Q$.

If $k=k(t)$ is not given by Eq. (\ref{req18a}), 
then $\lambda$ must be time dependent
(e.g., $\lambda \equiv \lambda(t)$).
Here one cannot directly perform the
MI analysis. However, still in this case, we have converted the explicit
spatial dependence into an explicit temporal dependence.
An important example of this type (the simplest one, in fact) is the case
of $k(t) \equiv k=$constant. Then,
\begin{eqnarray}
f(t) &=&-\frac{\sqrt{k}}{2} \tan \left( 2 \sqrt{k} (t+t^{\star}) \right)
\label{add2}
\\
\ell(t)&=&2 \ell(0) \frac{\cos \left(2 \sqrt{k} (t+t^{\star}) \right)}
{\cos \left(2 \sqrt{k} t^{\star} \right)} 
\label{add3}
\\
\tau(t)&=&-\frac{\cos^2 \left(2 \sqrt{k} t^{\star} \right)}{\ell(0)^2} \frac{f(t)}{k}
\label{add4}
\end{eqnarray} 
This case imposes a time-periodic driving term in 
Eq. (\ref{req15}) with frequency $4 \sqrt{k}$ which is nothing but the 
oscillation frequency naturally following from the Ehrenfest theorem.
In this viewpoint the phase divergence at $t_n=\pi (n+1/2)/(2 \sqrt{k})-t^\star$ 
(where $n=0,\pm 1, \pm 2, ...$) is  understandable. Indeed, in the 
case at hand, the ``chirp'' initial condition means existence of a current 
at the initial moment of time. Due to the quadratic potential this current 
periodically changes the direction (which is a straightforward consequence 
of  the Ehrenfest theorem). The change of the current direction is 
accompanied by the phase singularity.

From the above it is clear that the most interesting cases 
in the setting with the harmonic potential are the ones with
the inverse square dependence (of the trap amplitude, for a given $x$) 
on time of Eq. (\ref{req18a}),
as well as the one the regular harmonic trap of constant amplitude.
In both of these cases, as well as more generally, the lens transform
suggests the equivalence with a NLS equation with a gain. In these
special cases of interest the gain is constant or time periodic, 
suggesting that similar phenomenology to the one of the regular 
NLS may be present. A note of caution worth making here is that
in reality in this case, the evolution takes place in the setting
of Eq. (\ref{req1}), rather than that of Eq. (\ref{req15}). 
The motivation however is that upon suitable choice of the initial
condition (and for the types of traps discussed above), the two
equations (the GP and the NLS with the gain term)
are {\it equivalent at initial times}, hence one
may expect that the instability which is 
present in the latter will manifest itself in some manner in the former. 

However, to examine the details of the time evolution of this instability,
we perform numerical simulations of the Eq. (\ref{req1}) with appropriately
chosen modulationally stable as well as 
modulationally unstable initial conditions.


\section{Numerical Manifestations of the Modulational Instability for NLS
with Quadratic Potential}

\subsection{Time-Dependent Potential}

Perhaps the most interesting case (due to the suggested analogy with 
an NLS with a constant coefficient gain, where the modulational
stability analysis can be performed) is the case of 
$k(t) = (t+t^{*})^{-2}/16$, which we now examine numerically.

Notice that in our numerical investigations, we will apply a loss
term to Eq. (\ref{req1}) close to the boundaries to emulate the loss of
particles from the trap.

\begin{figure}
\centering
{\epsfig{file=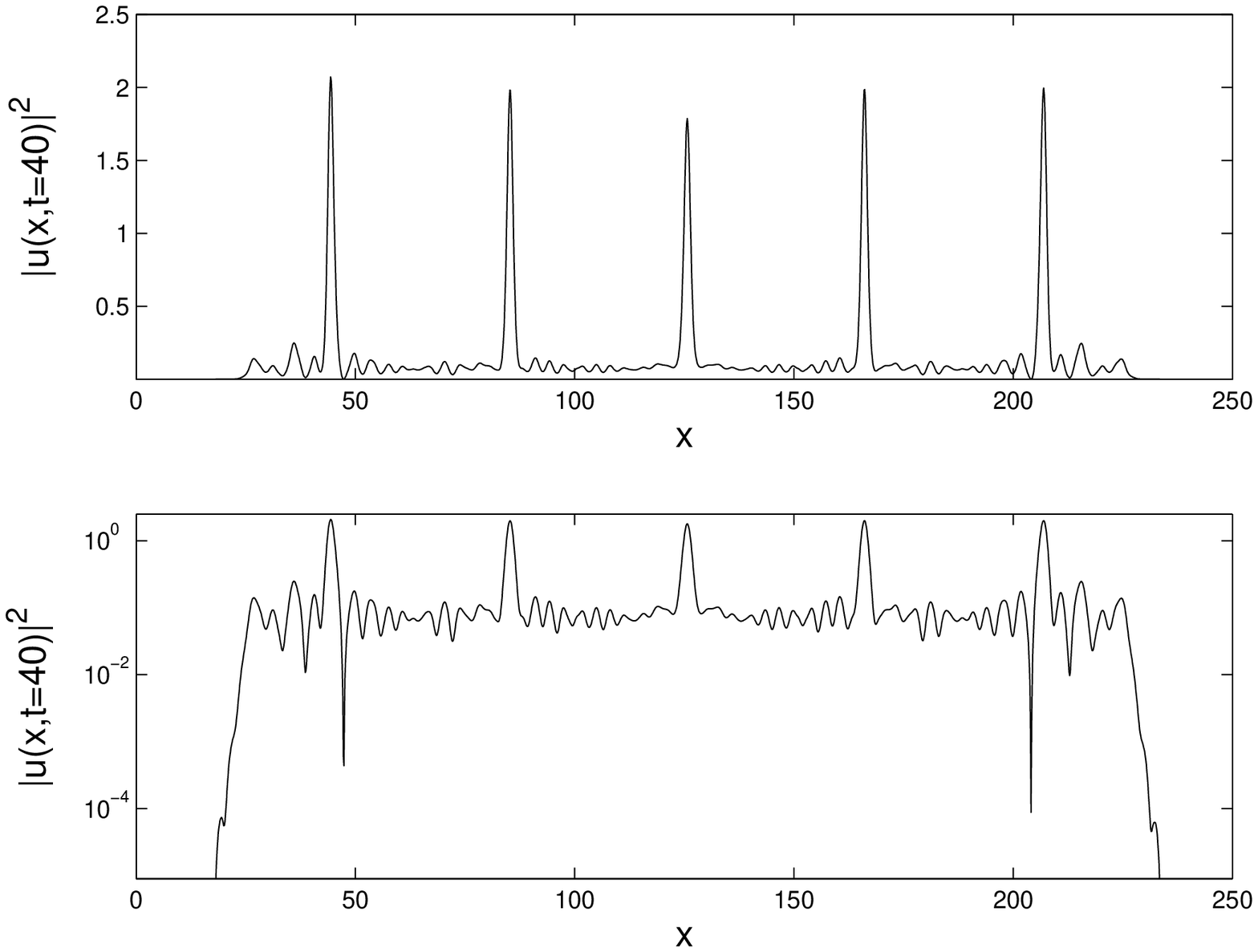, width=6.4cm,angle=0, clip=}}
{\epsfig{file=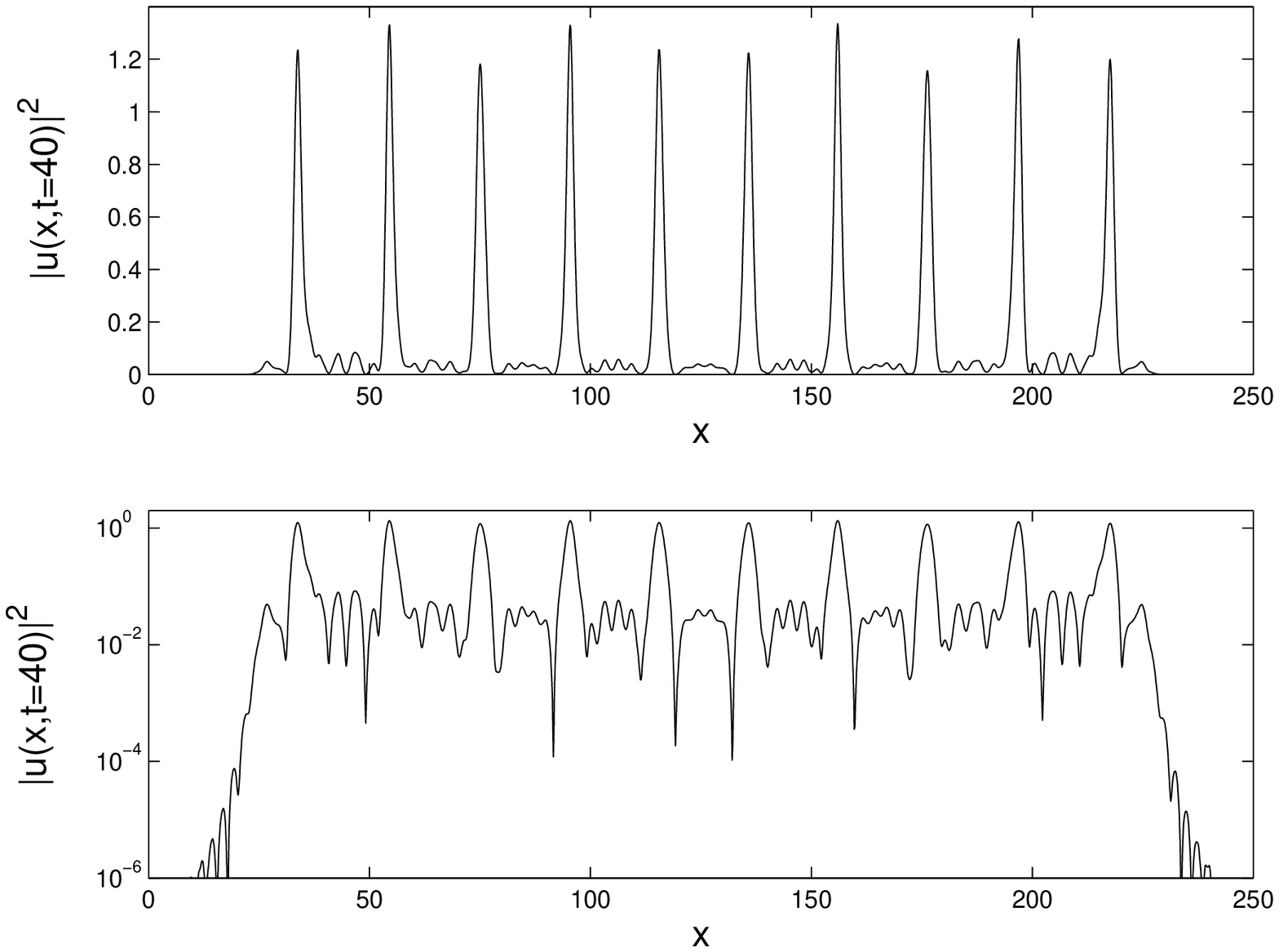, width=6.4cm,angle=0, clip=}}
{\epsfig{file=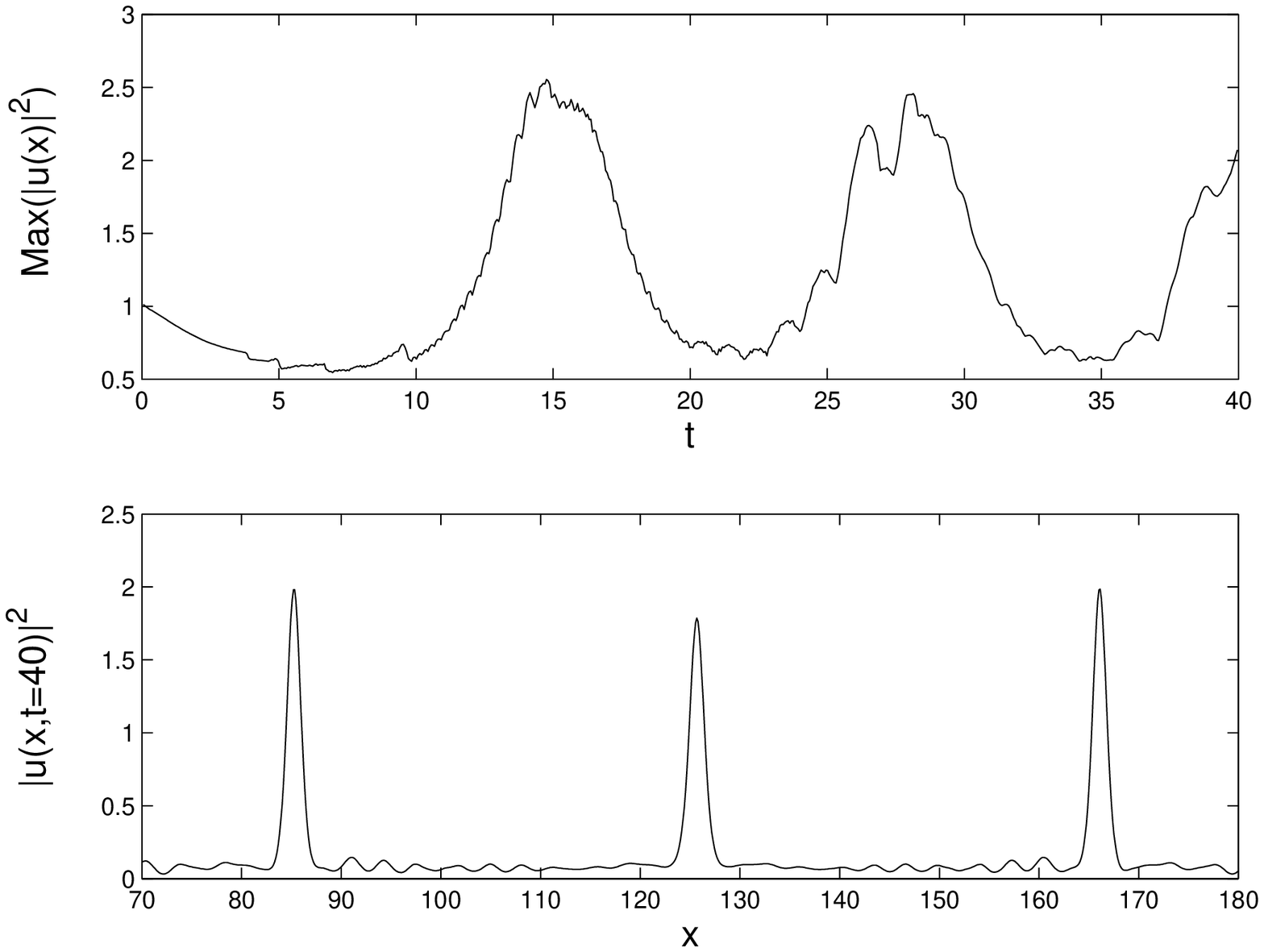, width=6.4cm,angle=0, clip=}}
{\epsfig{file=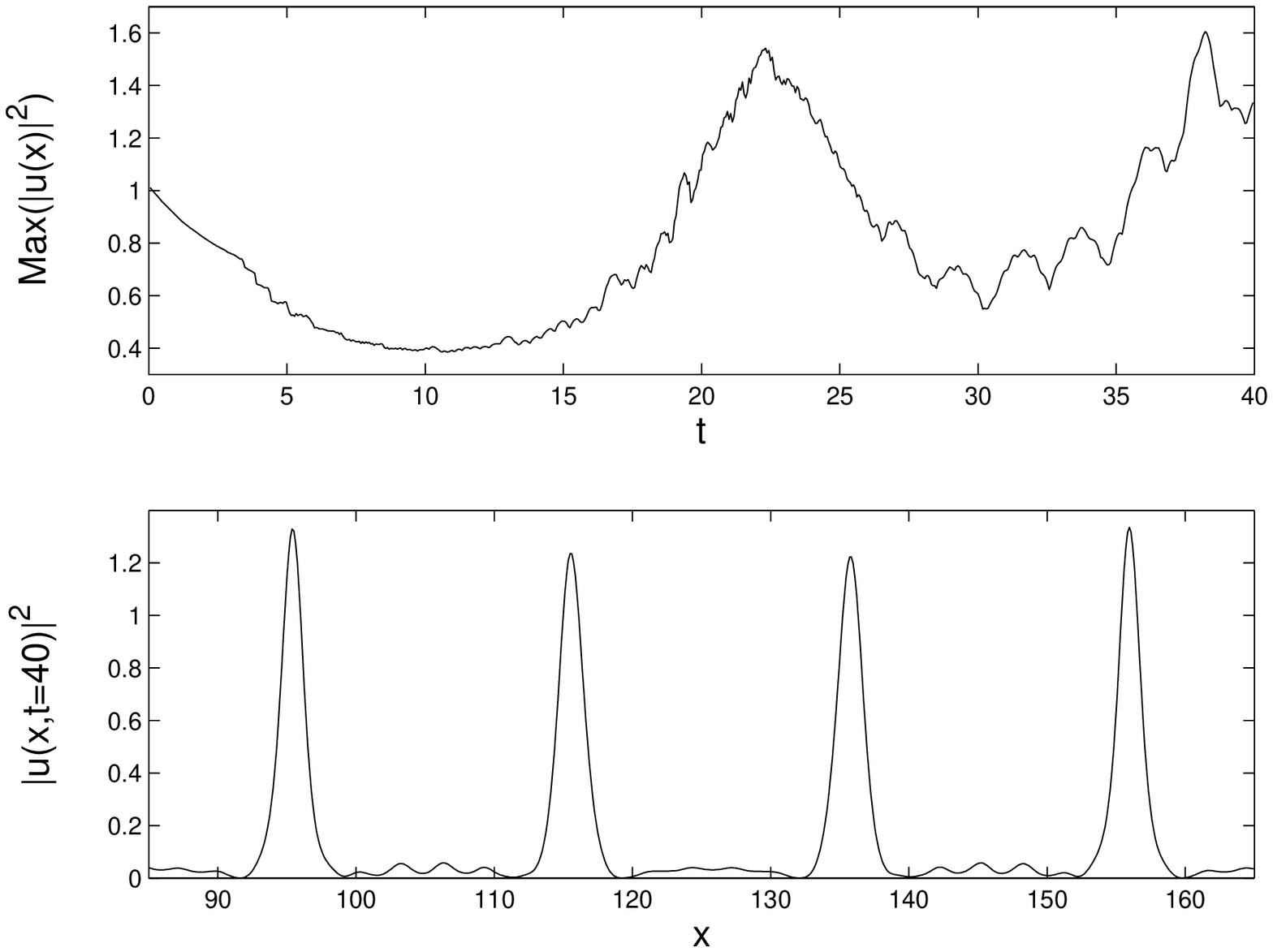, width=6.4cm,angle=0, clip=}}
\caption{The evolution of an initial condition of the form of 
Eq. (\ref{add6}) for the time-dependent potential of Eq. (\ref{req18a}).
The left panels show the case of $Q=1$, while the right ones the case of
$Q=2$. The panels show respectively: the profile of  $|u(x)|^2$ at $t=40$,
in the top panel; the same profile is shown 
in a semi-logarithmic plot in the second from the top
panel (clearly indicating the exponential nature of the localization).
The time evolution of the maximum amplitude of the configuration is
shown in the third (from the top) panel, while the bottom panel shows
a detail of the top one (indicating the clearly solitary structure of
the corresponding pulses).}
\label{rfig2}
\end{figure}

The first case that we studied in this setting was the one of an initial
condition
\begin{eqnarray}
u(x,t)= \exp(i  \frac{x^2}{8 t^{\star}} ) \left(1 + \epsilon \cos(Q x) 
\right), 
\label{add6}
\end{eqnarray}
suggested by Eqs. (\ref{req11}) and (\ref{req18a})-(\ref{req18c}).
$\epsilon$ was typically chosen in the range $0.01-0.1$ without significant
variation in the qualitative nature of the results. The parameter $t^{\star}$
was set to $1$. The results are shown in Fig. \ref{rfig2}, for the case of
$Q=1$ (left panels) and $Q=2$ (right panels).

\begin{figure}
\centering
{\epsfig{file=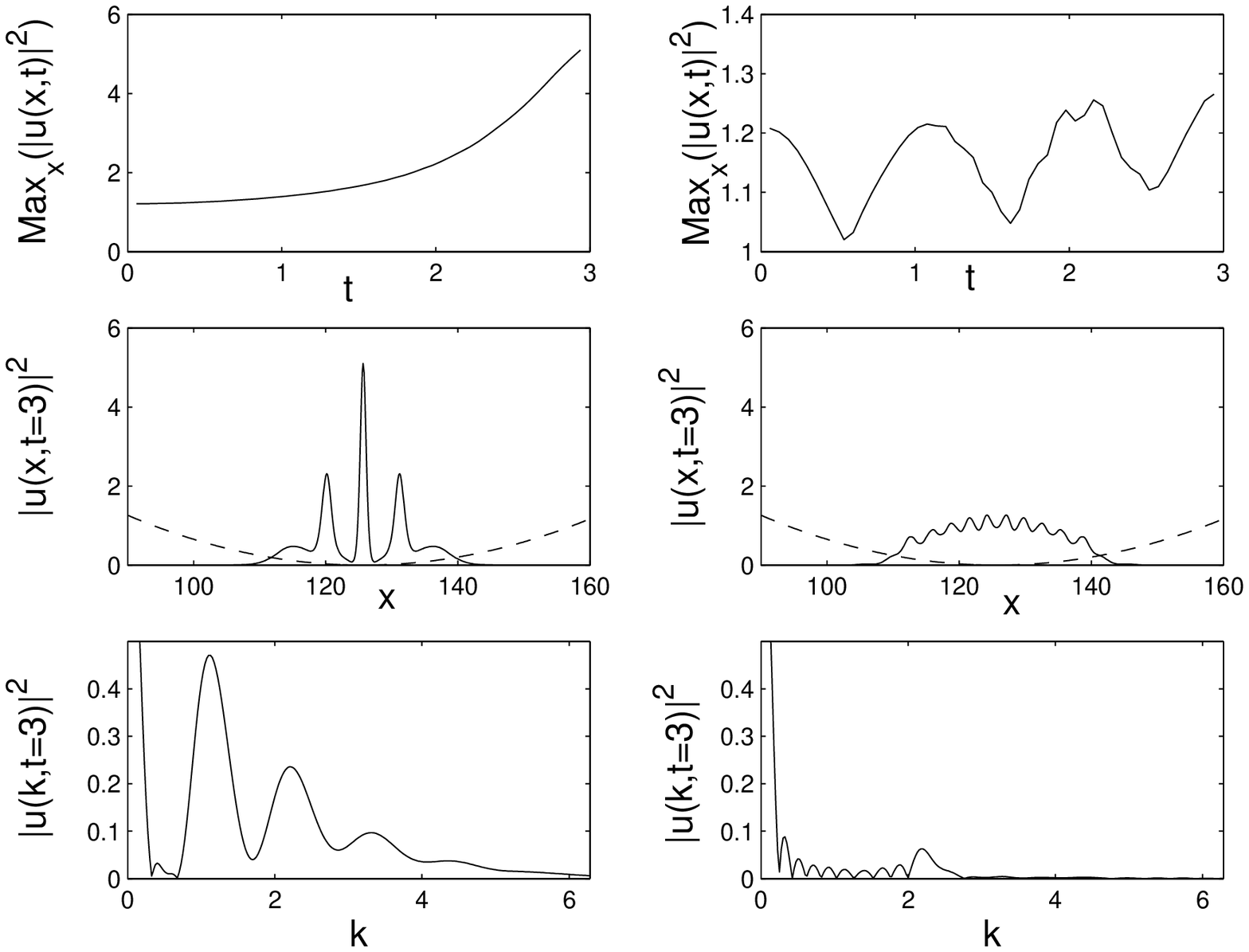, width=8.4cm,angle=0, clip=}}
\caption{The cases of $Q=1$ (left panels) and $Q=2$ (right panels)
are shown for the initial condition of Eq. (\ref{add7}). The top
panel shows the evolution of the maximum amplitude as a function of
time (for short times), the middle panel shows the mod-squared spatial profile
for $t=3$ (the dashed line here illustrates the trap at this time),
while the bottom panel shows Fourier transform for the same time ($t=3$).
$t^{\star}=5$ has been used here.}
\label{rfig4}
\end{figure}

It is clear from the time evolution shown in the figure that in this 
setting we obtain (and that is one of the main findings
of this work) a {\it soliton wave train}, formed as a result of
the instability, starting from such a modulated plane wave initial condition. 
An interesting feature of the obtained soliton train is that emerging 
solitons are of approximately equal shapes (amplitudes) in the presence 
of a broadening 
parabolic potential; in the case when the latter is static created solitons 
have essentially different shapes depending on their positions in space, 
see e.g., Fig.~\ref{rfig4}.

\begin{figure}
\centering
{\epsfig{file=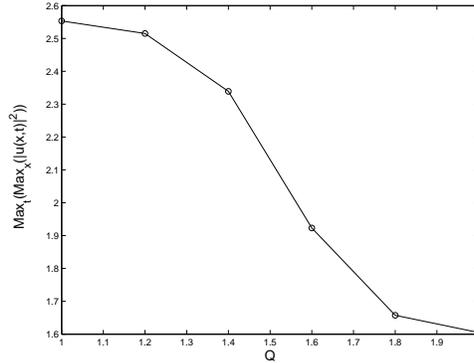, width=6.4cm,angle=0, clip=}}
\caption{The maximal amplitude (over space and time) for runs up to 
$t=40$, is shown 
for different values of the wavenumber $Q$ of the perturbation.}
\label{rfig3}
\end{figure}

One can argue that this outcome may not be a result of the modulational
instability given that both modulationally stable and unstable $Q$'s lead
to such a manifestation. 
However a careful inspection of the details of the evolution
(see also the short time runs reported below) outrules that possibility.
In particular the two features that happen for modulationally unstable
wavenumbers are:
\begin{itemize} 
\item The instability is manifested {\it at earlier times} (see in particular
the comparison of the third panels of Fig. \ref{rfig2}).
\item The instability leads to {\it larger amplitudes} in the modulationally
unstable regime (see e.g., the comparison of the fourth panels 
of Fig. \ref{rfig2}), than in the modulationally stable one. This is 
also clearly
shown in Fig. \ref{rfig3}, where the cases with different $Q$ in the
perturbation have been examined (for amplitude of the
original plane wave $\phi=1$), showing a clearly larger
amplitude tendency for ``unstable wavenumbers'' of $Q < \sqrt{2}$ in this
case.
\end{itemize} 

The reason why in practice the instability occurs in both cases is that
the dynamics of the potential in Eq. (\ref{req1}) mix the wavenumbers of
the original perturbation and eventually result in the excitation of 
modulationally unstable ones. However, this only happens later 
(because firstly the modulationally unstable $Q$'s need to be excited)
and with a smaller amplitude in this case.

We also tried a different initial condition motivated by the experimental
settings that led to the observation of bright matter wave solitons 
\cite{bright}. In particular, in these settings, a Feshbach resonance
is used to tune the sign of the nonlinearity (in the case of Eq. (\ref{req1})
the sign of $s$), starting from the repulsive case of $s<0$ and getting
to the attractive case of $s>0$, as time evolves in the experiment.
Given that in the case of $s<0$, the ground state of the system consists
(approximately) of the so-called Thomas-Fermi cloud \cite{review}, we
initialized the system in such a state, emulating the time (in the duration
of the experiment) in which the system is at $s<0$) and evolved the system
from such an initial condition. In this case $u(x,t=0)$ was of the form:
\begin{eqnarray}
u(x,t=0)=u_{TF} \left(1 + \epsilon \cos(Q x) \right).
\label{add7}
\end{eqnarray}
$u_{TF}=\sqrt{{\rm max}_x (0,\mu-V(x,t=0))}$ \cite{review}.
The chemical potential $\mu$ was chosen as $\mu=1$ in this case.

A particular example of this type for $\epsilon=0.1$, $Q=1$ (left 
panels) and $Q=2$ (right panels) is shown in Fig. \ref{rfig4}.

In this case, we only show short time dynamics, because at longer
times the Thomas-Fermi (which is not functionally close to the ground
state of the case with $s=1$) will be destroyed, leading to large
amplitude localized excitations independently of the initial
value of $Q$. In fact, this short time experiment illustrates all the
points that we made about (modulationally stable and unstable) short
time evolution previously. The modulationally unstable case of $Q=1$
rapidly develops the instability and deforms into a solitary wave train
pattern. On the contrary, for the short times reported in Fig. \ref{rfig4},
the modulationally stable case is limited to benign oscillations of small
amplitude. In the case of $Q=1$, the sidebands clearly form, indicating the
manifestation of the MI. However, notice also, as highlighted previously,
that in the case of $Q=2$, the dynamics of Eq. (\ref{req1}) eventually
tends to excite modulationally unstable wavenumbers and hence will 
also result (for longer times) in localization. 

\subsection{Time-Independent Potential}

In the case in which the potential is time independent
we first (once again) tried an initial condition with a modulation added to 
the plane wave in the form
\begin{eqnarray}
u = 1+ \epsilon \cos(Qx)
\label{add8}
\end{eqnarray}
Notice that in this case the chirp was not used in the initial
condition as it does not rid the equation of the explicit temporal dependence.

\begin{figure}
\centering
{\epsfig{file=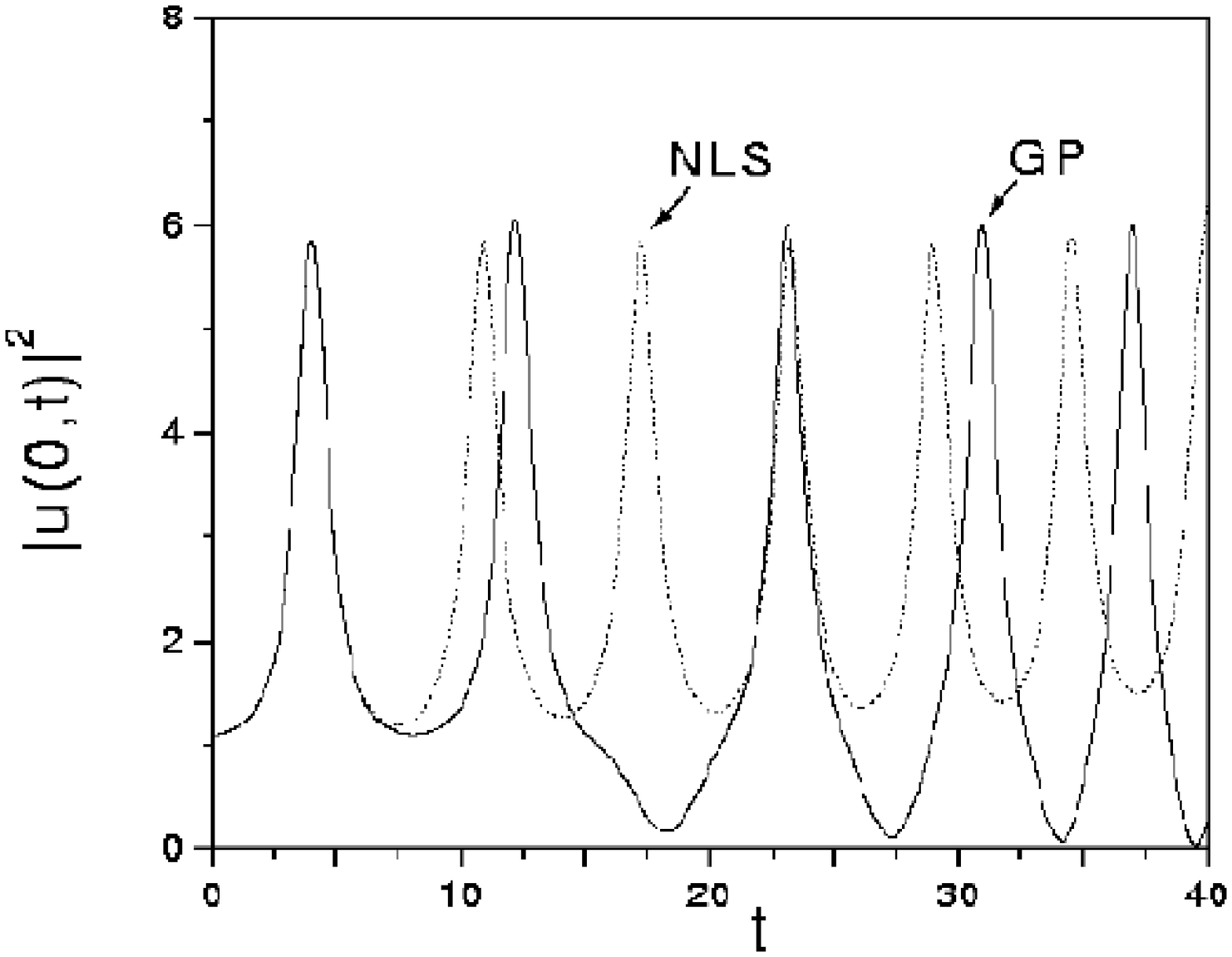, width=6.4cm,angle=270, clip=}}
{\epsfig{file=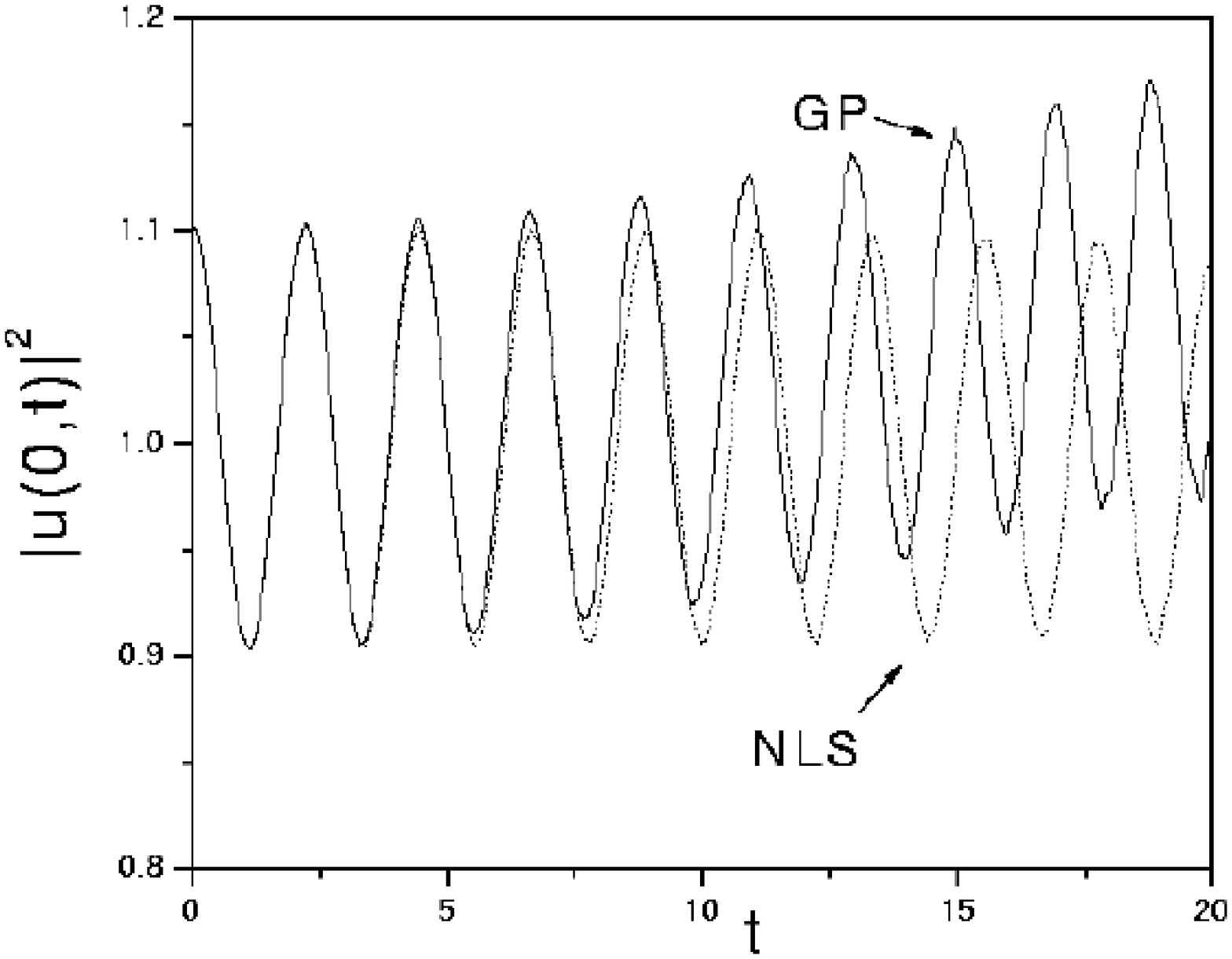, width=6.4cm,angle=270, clip=}}
\caption{The time evolution of the amplitude at $x=0$ ($|u(0,t)|^2$) is shown in
the left panel for $Q=1$ and in the right one for $Q=2$. The comparison
of the GP Equation (solid line) with the corresponding case for the NLS 
(dotted line) is also illustrated.}
\label{rfig5}
\end{figure}

In this case the findings, once again for $Q=1$ and $Q=2$, are shown in
Fig. \ref{rfig5}. $\epsilon=0.05$ was used in Eq. (\ref{add8}); $k=0.0001$.

In both cases, for the GP equation, due to the presence of the potential,
the condensate will become peaked towards the center, gradually as
time evolves. However, the development of the instability is clear
from the comparison of the corresponding amplitudes of the oscillation
of the norm field as a function of time. 

\begin{figure}
\centering
{\epsfig{file=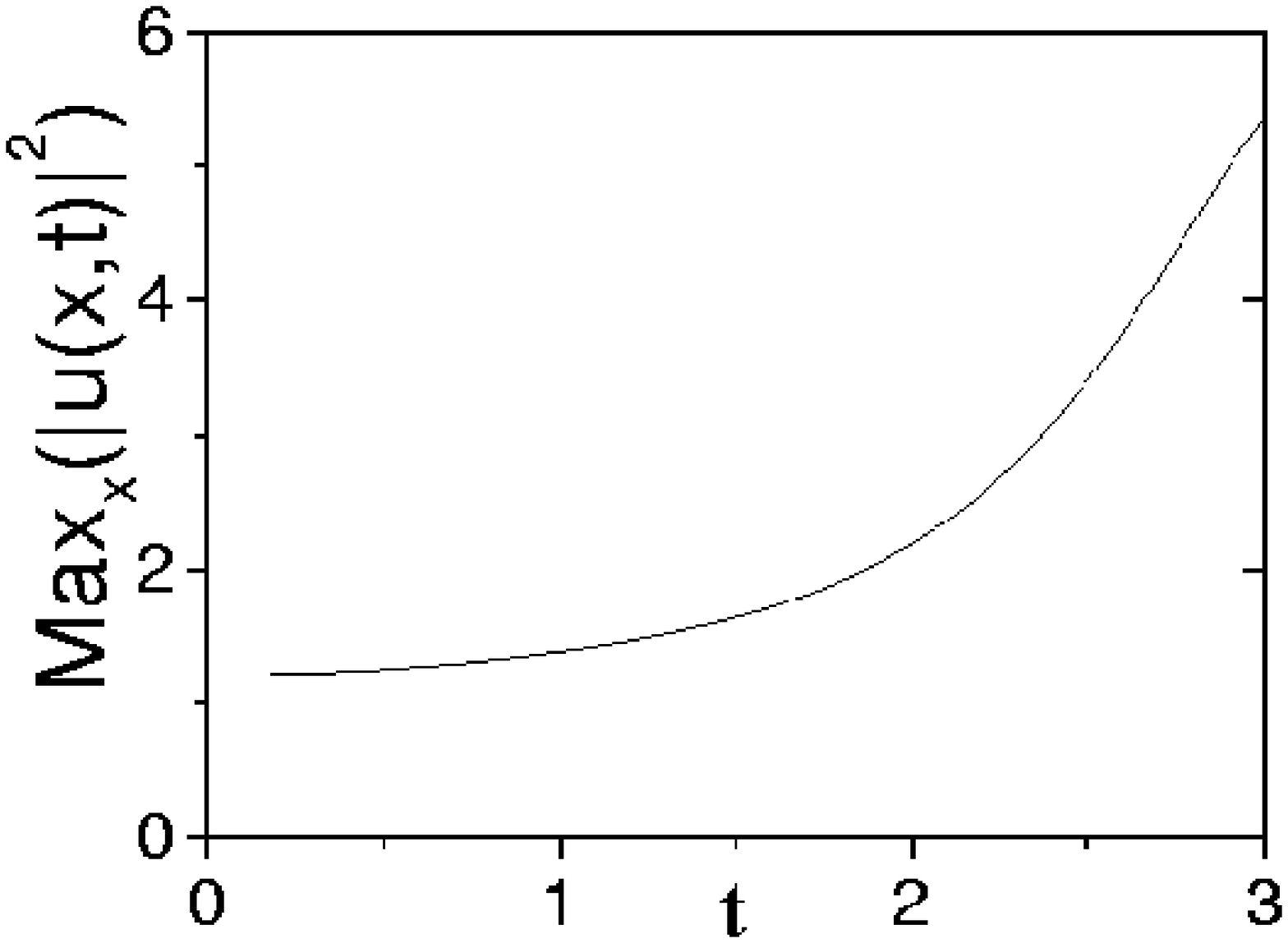, width=6.4cm,angle=0, clip=}}
{\epsfig{file=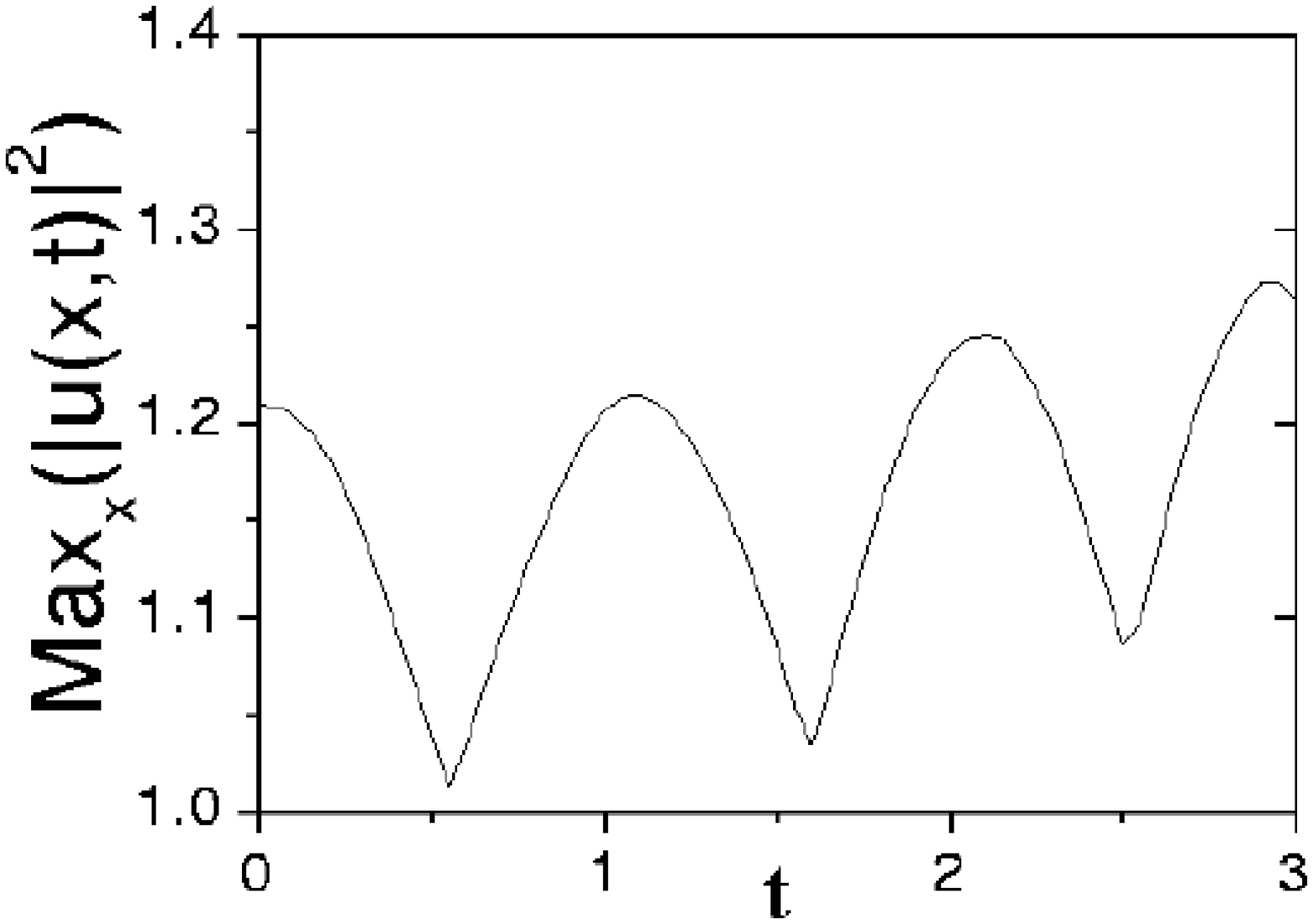, width=6.4cm,angle=0, clip=}}
{\epsfig{file=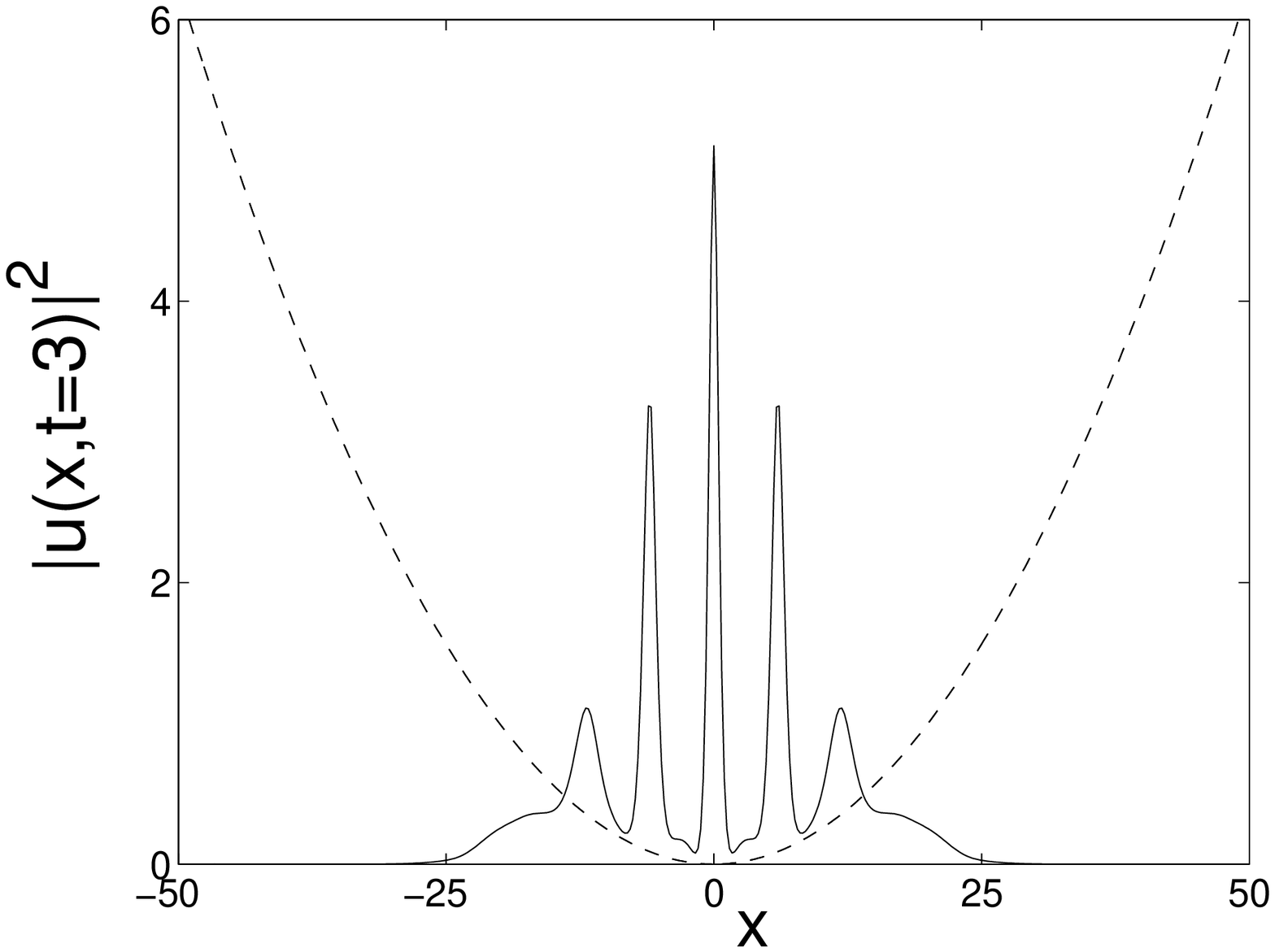, width=6.4cm,angle=0, clip=}}
{\epsfig{file=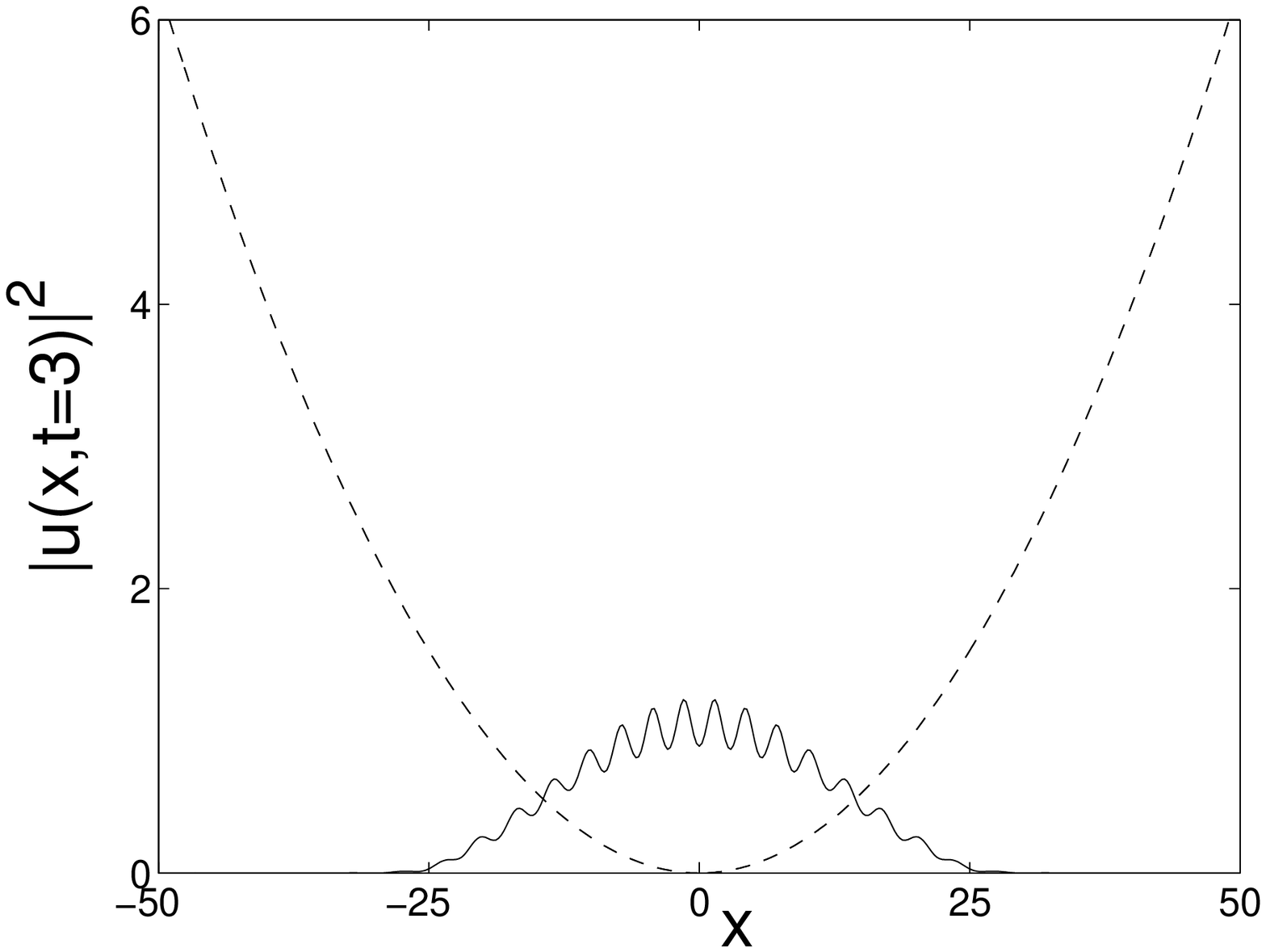, width=6.4cm,angle=0, clip=}}
{\epsfig{file=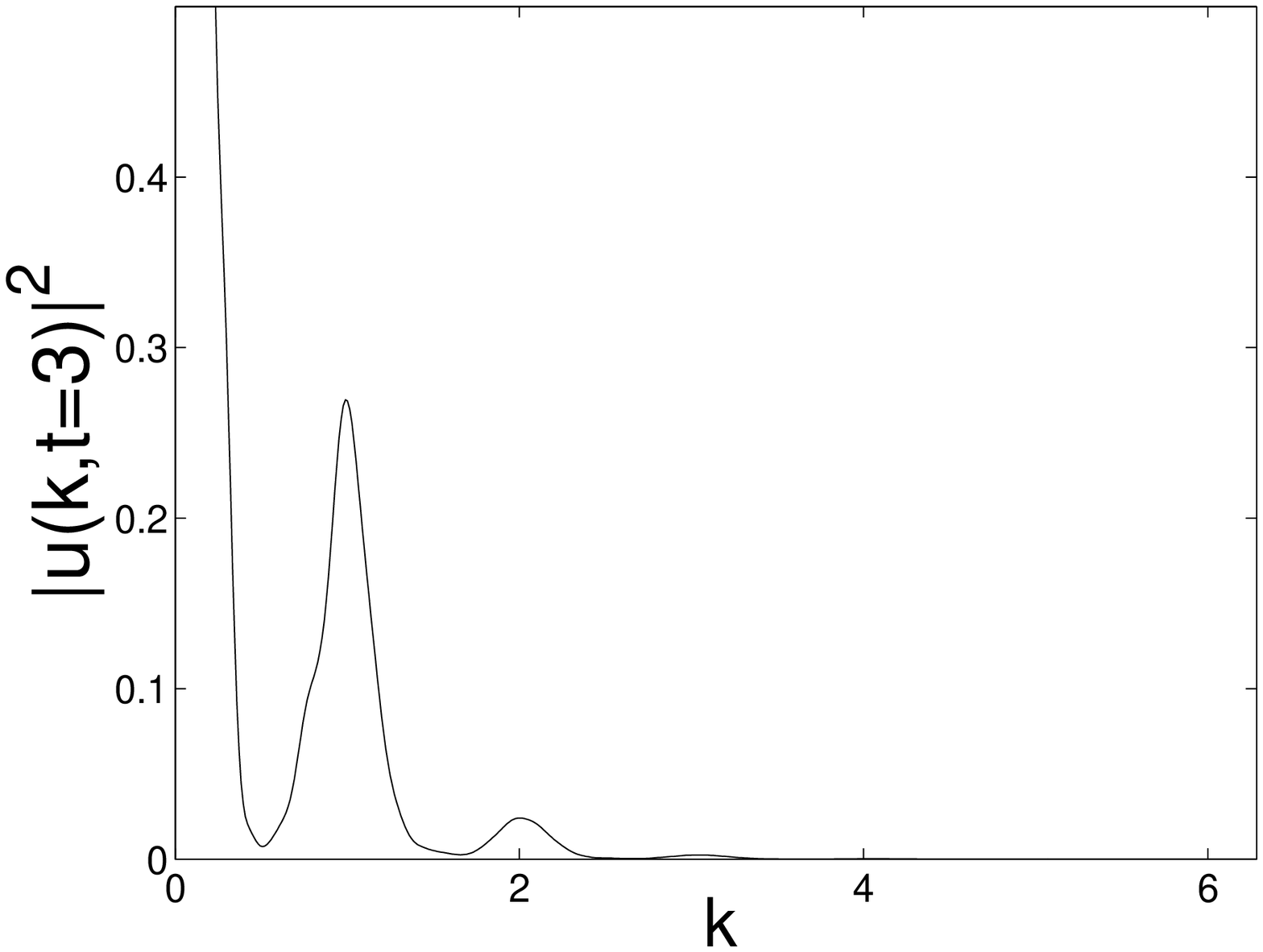, width=6.4cm,angle=0, clip=}}
{\epsfig{file=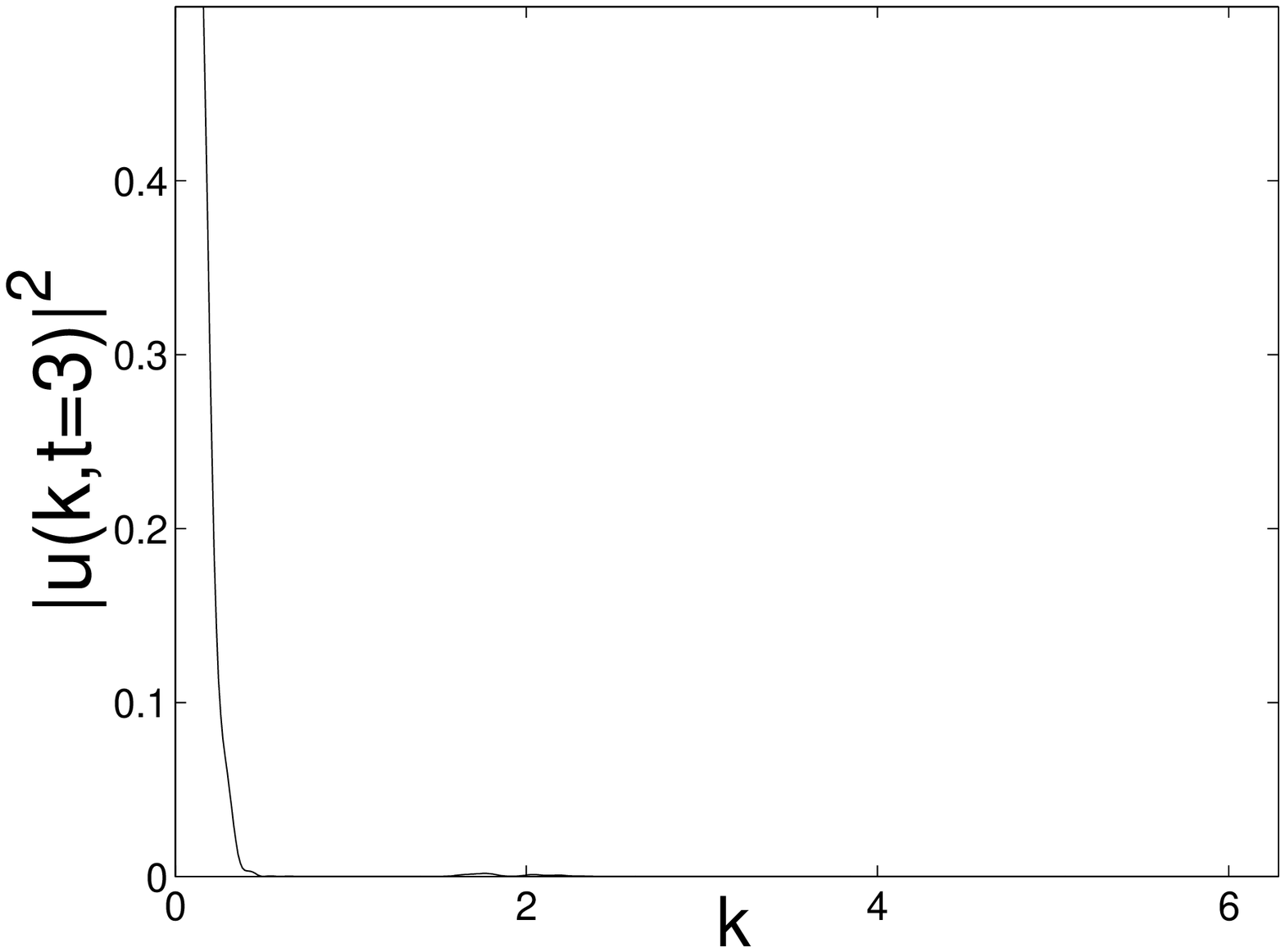, width=6.4cm,angle=0, clip=}}
\caption{The same as Fig. \ref{rfig4}, but for the case of 
$k(t)=0.0025=$constant. The left panels correspond to $Q=1$,
and the right ones to $Q=2$. }
\label{rfig6}
\end{figure}

The case with the Thomas-Fermi initial condition of Eq. (\ref{add7})
is shown in Fig. \ref{rfig6}. $k=0.0025$ in this case and once again
the cases of $Q=1$ and $Q=2$ are shown in the left and right
panels respectively.

The development of the instability for short times is once again
clear for the modulationally unstable case of $Q=1$, leading to
the formation of a wave train, while in the modulationally stable
case, the perturbation is not amplified. For longer times, the
destruction of the TF cloud will eventually lead in both cases
to the generation of very strongly localized patterns. However,
in essence here, we take advantage of the separation of time
scales for the appearance of the MI and for the destruction of the
TF, to illustrate in the short time evolution, the development of the 
former instability.

\section{Conclusions}

In this work, we have examined the problem of modulational
instabilities of plane waves in the context of Gross-Pitaevskii equations
with an external (in particular quadratic) potential. The motivation
for this study was its direct link 
to current experimental realizations of Bose-Einstein condensates.

A lens transformation was
used to cast the problem in a rescaled space and time frame (in a
way very reminiscent of the scaling in problems related to focusing
\cite{sulem,skk}). In this rescaled frame, the external potential
can be viewed as a form of external growth. For specific forms
of temporal dependence of the prefactor of the harmonic potential (e.g., 
for $k(t) = (t+t^{*})^{-2}/16$), the resulting
growth term is constant. In such a context once again the MI analysis
can be carried through completely, producing similar
conditions, but now in the new dynamically rescaled frame/variables
(which can be appropriately re-cast in the original variables).
This singles out the case of a temporally dependent potential
of inverse square dependence with time. Another case which was
also examined due to its direct relevance to the experiment
was the one of the constant amplitude trap.

Both of these cases were analyzed theoretically and then studied
in detail numerically. The theoretical predictions for modulational
instability were verified by the numerical experiments. This
was most clearly identified
for short time dynamical evolution results
that permit to clearly identify the instability
through the formation of localized pulses and trains thereof.
For longer times, trains are also formed for modulationally stable
cases (due to the eventual excitation in the dynamics of unstable
wavenumbers). However, there are still ``stronger'' signatures
of the instability in the unstable cases (such as the larger amplitude
of the resulting excitations in such cases). 

The main aim of this work is to advocate the use of the MI
as an experimental tool to generate solitonic trains in Bose-Einstein
condensates. Our theoretical investigation and numerical findings
clearly support the 
formation of such trains in the context of the GP equation initialized
with an appropriate modulation and possibly a chirp. The latter
is needed in the case of the time-dependent trap that we have examined
herein and which we argue may also be interesting to try to create in 
experimental settings. Let us note in passing that traps with this type
of time dependence of their amplitude have also been suggested as being
of interest in quite different setups such as the study of 
explosion/implosion dualities for the quintic (critical) GP \cite{ghosh}.
However, our findings should be observable
{\it even without} the time-dependent trap, as we have demonstrated.
The appropriate modulation in the condensate initial condition can
be generated by placing the condensate in an optical lattice \cite{lattice},
while the chirp can also be produced using appropriate phase engineering
techniques which are currently experimentally available; see e.g.,
\cite{dexp2}. We believe that such experiments are within the 
realm of present experimental capabilities and hope that these theoretical
findings may motivate their realization in the near future.

PGK gratefully acknowledges support from a University of Massachusetts
Faculty Research Grant, from the Clay Foundation through a Special 
Project Prize Fellowship and from the NSF through DMS-0204585.
VVK gratefully acknowledges support 
from the European grant COSYC n.o {HPRN-CT}-2000-00158.


\begin{thebibliography}{0}

\bibitem{review}  F. Dalfovo, S. Giorgini, L. P. Pitaevskii, and S.
Stringari, 
Rev. Mod. Phys. {\bf 71}, 463
(1999);
A.J. Leggett, Rev. Mod. Phys. {\bf 73}, 307 (2001).

\bibitem{vortex}  M.R. Matthews {\it et al.}, Phys. Rev. Lett. {\bf 83},
2498 (1999); K.W. Madison {\it et al.} Phys. Rev. Lett. {\bf 84}, 806
(2000); S. Inouye {\em et al.}, Phys. Rev. Lett.{\bf {87}}, 080402 (2001).

\bibitem{vl}  J.R. Abo-Shaeer {\it et al}, Science {\bf 292}, 476 (2001);
J.R. Abo-Shaeer, C. Raman and W. Ketterle, Phys. Rev. Lett. {\bf {88}},
070409 (2002); P. Engels {\it et al}, Phys. Rev. Lett. {\bf 89}, 100403
(2002).

\bibitem{dark}  S. Burger {\it et al.}, Phys. Rev. Lett. {\bf 83}, 5198
(1999). 

\bibitem{dexp2}  J. Denschlag {\it et al.}, Science {\bf 287}, 97 (2000).

\bibitem{nsbec}  B.P. Anderson {\it et al.}, Phys. Rev. Lett. {\bf 86}, 2926
(2001).

\bibitem{bright}  K.E. Strecker {\em et al.}, Nature {\bf 417}, 150 (2002);
L. Khaykovich {\em et al.}, Science {\bf 296}, 1290 (2002).

\bibitem{sulem} C. Sulem and P.L. Sulem,
\newblock {\it The Nonlinear Schr{\"o}dinger Equation},
Springer-Verlag (New York, 1999).


\bibitem{benjamin67} T.B. Benjamin and J.E. Feir, 
J. Fluid. Mech. {\bf 27}, 417 (1967).

\bibitem{ostrovskii69} L.A. Ostrovskii, 
Sov. Phys. JETP {\bf 24}, 797 (1969).

\bibitem{taniuti68} T. Taniuti and H. Washimi, 
Phys. Rev. Lett. {\bf 21}, 209 (1968); A. Hasegawa, 
Phys. Rev. Lett. {\bf 24}, 1165 (1970).

\bibitem{konotop} V.V. Konotop and M. Salerno,
\newblock Phys. Rev. A {\bf 65}, 021602(R) (2002).

\bibitem{smerzi}  A. Smerzi, A. Trombettoni, P.G. Kevrekidis,
and  A.R. Bishop, Phys. Rev. Lett., {\bf 89}, 170402 (2002).

\bibitem{cattal} F.S. Cataliotti {\it et al.}, 
cond-mat/0207139.

\bibitem{kasevich} M. Kasevich and A. Tuchman 
(private communication).


\bibitem{book} A. Hasegawa and W.F. Brinkman, IEEE J. Quantum Electron. 16, 694
(1980); K. Tai, A. Tomita, and A. Hasegawa, Phys. Rev. Lett. 56, 135 (1986).


\bibitem{skk} see e.g., C.I. Siettos, I.G. Kevrekidis and
P.G. Kevrekidis, nlin.PS/0204030 and references therein
(Nonlinearity, in press 2003).

\bibitem{ghosh} P.K. Ghosh, cond-mat/0109073.

\bibitem{lattice} see e.g., F.S. Cataliotti, S. Burger, C. Fort, P. Maddaloni,
F. Minardi, A. Trombettoni, A. Smerzi, and M. Inguscio,
Science {\bf 293}, 843 (2001); M. Greiner, O. Mandel, T. Esslinger, 
T.W. H\"{a}nsch, and I. Bloch, Nature {\bf 415}, 39 (2002).

 




\end{thebibliography}
\end{document}